\documentclass[twoside,12pt]{article}

\usepackage[dvips]{graphics,epsfig}
\usepackage[latin1]{inputenc}
\usepackage{setspace}
\usepackage{amsfonts}

\usepackage{textcomp}
\usepackage{amsmath}
\usepackage{graphicx}
\usepackage{epstopdf}

\usepackage{subfig}

\def\laq{\raise 0.4 ex \hbox{$<$}\kern -0.8 em\lower 0.62 ex\hbox{$\sim$}}
\def\gaq{\raise 0.4 ex \hbox{$>$}\kern -0.7 em\lower 0.62 ex\hbox{$\sim$}}

\def\beq{\begin{equation}}
\def\eeq{\end{equation}}
\def\beqa{\begin{eqnarray}} 
\def\eeqa{\end{eqnarray}}

\begin{document}
\pagestyle{plain}

\begin{flushright}
{\bf DRAFT VERSION}\\
Out 15, 2019
\end{flushright}
\vspace{15mm}

\begin{center}

{\Large\bf Multiplicative Noise in Euclidean Schwarzschild Manifold}

\vspace*{1.0cm}

Matheus S. Soares$^{*}$, Nami F. Svaiter$^{**}$\\
\vspace*{0.5cm}
{Centro Brasileiro de Pesquisas F\'{\i}sicas\\
Rua Xavier Sigaud, 150 - Urca, Rio de Janeiro - RJ, 22290-180, Brazil}\\

\vspace*{1.0cm}

Carlos A. D. Zarro$^{***}$\\
\vspace*{0.5cm}
{Instituto de F\'isica\\
 Universidade Federal do Rio de Janeiro,\\
Av. Athos da Silveira Ramos, 149 - Cidade Universit\'aria da Universidade Federal do Rio de Janeiro, Rio de Janeiro - RJ, 21941-909, Brazil}\\

\vspace*{2.0cm}
\end{center}

\begin{abstract}
We discuss a $\lambda\varphi^{4}+\rho\varphi^{6}$ scalar field model defined in the Euclidean section of the Schwarzschild solution of the Einstein equations in the presence of multiplicative noise. The multiplicative random noise is a model for fluctuations of the Hawking temperature. We adopt the standard procedure of averaging the noise dependent generating functional of connected correlation functions of the model. The dominant contribution to this quantity is represented by a series of the moments of the generating functional of correlation functions of the system. Positive and negative effective coupling constants appear in these integer moments. Fluctuations in the Hawking temperature are able to generate first-order phase transitions. Using the Gaussian approximation, we compute $\langle\varphi^{2}\rangle$ for arbitrary values of the strength of the noise. Due to the presence of the multiplicative noise, we show that  $\langle\varphi^{2}\rangle$ near the horizon must be written as a series of the the renormalized two-point correlation functions associated to a free scalar field in Euclidean Rindler manifold. 
\end{abstract}

\vfill
\noindent\underline{\hskip 140pt}\\[4pt]
{$^{*}$ E-mail address: matheus.soares@cbpf.br} \\
\noindent
{$^{**}$ E-mail address: nfuxsvai@cbpf.br} \\
\noindent
{$^{***}$ E-mail address: carlos.zarro@if.ufrj.br}

\newpage



\section{Introduction}\label{intro}

In this paper we investigate phenomena described by a semi-classical theory of gravity, where quantized matter fields are defined in a classical space-time. One of the most interesting prediction in this scenario is that a spherical uncharged non-rotating black hole emits thermal radiation. In a field theory framework, the above discussed result indicates a deep connection between geometric event horizons and thermal effects \cite{bek,bek2,hawk,candelas,fulling}. In the extended Schwarzschild solution of the Einstein equations it can be shown that the gravitational entropy is proportional to the area of the event horizon. With the advent of the gravitational wave astronomy, the semi-classical gravity paradigm can be tested \cite{corda,maria3}. 

This work is motivated by two points that have been discussed by many authors. The first one is that a deeper understanding of the black hole thermodynamics and the effects of the geometric event horizon may be achieved introducing the idea of a quantum open systems \cite{davies, hu}. The second one is that is necessary to investigate the effects of fluctuations of the Hawking temperature  \cite{flu1,flu2}.  
To overcome technical difficulties connected with the fluctuations of the black hole mass, here we use a quite simple and useful approach. First, to study open systems the literature have been used
differential equations with random coefficients. Also in the $\lambda\varphi^{4}$ scalar model at finite temperature, the coefficient of $\varphi^{2}$ is a function of the temperature. Therefore,  
the simplest assumption we can make to model fluctuation of the Hawking temperature is to add a random contribution to this coefficient. This
random coefficient can be considered as a local fluctuation of the Hawking temperature. For instance, the pseudo Riemannian Schwarzschild solution of the Einstein equations, after a Wick rotation becomes a Riemannian metric for $r>2M$, where the singularity at $r=2M$ is removed if the Euclidean time is periodic with period $8\pi M$. This periodicity in imaginary time defines a canonical ensemble at temperature $T=1/8\pi M$ \cite{ha1,pe, hawki}. Here, we introduce multiplicative noise in a scalar field model defined in the Euclidean section of the Schwarzschild solution. We use the term multiplicative noise to the situation where the noise couples quadratically with the scalar field.  It is important to point out that our approach is quite different from stochastic semi-classical gravity, where the induced fluctuations of the gravitational field can be obtained studying the Einstein Langevin equations. In this approach it is possible to obtain a self-consistent study of horizon fluctuations \cite{ech1,ech2}.
 
The aim of this paper is  to discuss connections between gravitational physics, statistical mechanics and statistical field theory. We study the $\lambda\varphi^{4}+\rho\varphi^{6}$ scalar model defined in the Euclidean section of the Schwarzschild solution of the Einstein equations in the presence of noise.  We are interested to use the approach that have been used in the theory of nonlinear dynamical systems with the presence of multiplicative noise \cite{noise, toral1, toral2}. The main characteristic of these equations is the introduction of a source of disorder or random noise in models describing macroscopic phenomena. A fundamental question is how the noise couples with the dynamical variables. On general grounds, the models are usually defined with an additive internal noise or a multiplicative external noise term. The origin of internal noise is an average effect of microscopic degree of freedom that are not otherwise taken into account.  On the other hand in the scenario discussing noise in spatially extended systems, a multiplicative noise is an external fluctuating parameter. As it has been discussed in the literature, the presence of multiplicative noise has interesting consequences, as for example noise-induced phase transitions \cite{bookojalvo,lefever}. In this work, we discuss phase transitions generated by multiplicative noise coupled with a scalar field defined in Euclidean black hole background. This multiplicative noise model is the random mass $d$-dimensional Landau-Ginzburg model on a black hole background. In the $\lambda\varphi^{4}$ scalar model, the renormalized squared mass is a regular function of the temperature. The behavior of scalar field models at finite temperature has been widely discussed in the literature \cite{bernard,Dolan,bookkapusta,ananos}. In a generic $d$-dimensional Euclidean space in the one loop approximation one can show that the renormalized squared mass is a sum of two terms: one correspond to zero temperature and other explicitly temperature dependent $i.e.$ $m^2(\beta) = m_{0}^2 + \delta m_{0}^2(\beta)$. Therefore, as it was discussed above, a random contribution $\delta m_{0}^{2}(x)$ added to $m_{0}^{2}$ can be  considered as a local perturbation of the Hawking temperature. This approach was also used by Dotsenko where a random temperature Landau-Ginzburg model defined in a $d$-dimensional Euclidean space was investigated \cite{Vik}. Following these ideas, we introduce a zero mean Gaussian white noise in the $\lambda\varphi^{4}+\rho\,\varphi^{6}$ scalar field model defined in Euclidean black hole background. The $\varphi(x)^{6}$ contribution  must be introduced to obtain a stable theory. We would like to stress that in principle one must include a random contribution to the time derivative kinetic term following the scenario discussed in  \cite{arias}. As we discussed we are studying  quantized matter fields in a classical space-time. Another argument to avoid this random contribution is the following: the mass introduces a characteristic length, therefore our description of the fluctuations is valid for distances larger than this characteristic length. By this reason, it is not necessary to introduce randomness in the time derivative operator.  

In the scenario of continuous phase transitions in equilibrium statistical mechanics, the usual approach is to use functional methods. To implement a functional approach for scalar fields and noise in the Euclidean section of the Schwarzschild solution of the Einstein equations, it is necessary to define a functional integral of the system  using the scalar action with the presence of noise. From the  generating functional of correlation functions for one realization of the noise of the model, we define the generating functional of connected correlation functions also for one realization of the noise. In order to obtain the noise averaged generating functional of the connected correlation functions of the model, we use the distributional zeta-function method \cite{distributional,distributional2, zarro, robinson2}. At this point, with these mathematical tools, it is worth emphasizing how it is possible to go one step further starting from the Fawcett and Hawking results \cite{fawcett,hawkinginteraction}. These authors discussed the behavior of a scalar field near a evaporating black hole. A scalar field in the ordered phase can go from a spontaneous broken symmetry phase to the disordered phase near a hot black hole, since at sufficiently high temperature a spontaneous broken symmetry can be restored. Now, suppose a very high temperature black hole and the scalar field in the disordered phase. Taking into account fluctuations in the Hawking temperature, it appears transitions from a disordered phase to an ordered ones with decay of metastable phases with first-ordered phase transitions.

For instance, random noise was introduced in space-times with event horizons in a quite different situation. In   \cite{krein, bessa} it was investigated the influence of fluctuations in the event horizon on the transition rate of a two-level system interacting with a quantum field in Rindler space. The main result of these works is that in the case of a scalar field interacting with a Unruh-Dewitt detector, the correction to the response function has a Fermi-Dirac factor. For the case of massless Dirac field coupled to a detector, i.e., a fermionic monopole moment operator, the spectral density of fermion field is also modified due to the horizon fluctuations, where it appears a Bose-Einstein contribution. 

The organization of this paper is the following. In Sec. \ref{sec:random} we discuss the Euclidean section of the Schwarzschild solution of the Einstein equations. In Sec. \ref{sec:vacuum} we discuss the structure of the fields in each integer moment of the generating functional of all correlation functions. In Sec. \ref{sec:instantons},  we compute $\langle\varphi^{2}\rangle$ in the weak- and strong-noise situations. In Sec. \ref{sec:Rindler} to find the $\langle\varphi^{2}\rangle$ in the presence of multiplicative noise near the horizon, we discuss the two-point correlation function in Euclidean Rindler manifold. Conclusions are given Sec. \ref{conclusions}.  We assume that $\hbar=c=k_{B}=G=1$.

\section{Interacting scalar field in Euclidean section of the\\ Schwarzschild solution}\label{sec:random}

In classical statistical mechanics of Hamiltonian systems, any state is a probability measure on the phase space. The expectation value of any observable can be obtained from an average constructed with the Gibbs measure. For systems described in the continuum with infinitely many degrees of freedom, this framework can be maintained. For instance, Euclidean functional methods with functionals and probability measures, introduced classical probabilistic concepts in quantum field theory \cite{drouffe}. The Euclidean correlation functions, i.e., the Schwinger functions, are the analytic continuation of vacuum expectation values for imaginary time of the Wightman functions \cite{schwinger,nakano,symanzik2, symanzik}. For a scalar field, these $n$-point correlation functions, which are the moments of probability measure, are defined by
\begin{equation}\label{eq:correlationfunction}
\langle\varphi(x_{1})..\varphi(x_{k})\rangle=\frac{1}{Z}\int [d\varphi]\prod_{i=1}^{k}\varphi(x_{i}) \exp\left(-S(\varphi)\right), 
\end{equation}
where $[d\varphi]$ is a functional measure, i.e., a measure in the space of all field configurations, given by $[d\varphi]=\prod_{x} d\varphi(x)$ and $S(\varphi)$ is the Euclidean action of the system.  To discuss non-trivial effects generated by random noise on a Euclidean black hole background, we must briefly present the Schwarzschild solution.

The pseudo Riemannian Schwarzschild metric is defined by the line element 
\begin{equation}\label{eq:line element1}
ds^2= -\left(1-\frac{2M}{r}\right)dt^2+\left(1-\frac{2M}{r}\right)^{-1}dr^2 +r^2(d\theta^{2}+\sin^{2}\theta d\phi^{2}).
\end{equation}
For $\tau=it$ we obtain a positive Euclidean metric for $r>2M$. Introducing the radial coordinate 
\begin{equation}
x=4M\sqrt{\left(1-\frac{2M}{r}\right)},
\end{equation}
the line element becomes
\begin{equation}\label{eq:line element2}
ds^2=\frac{x^2}{16M^2}d\tau^2+\frac{r^4}{16M^4}dx^2
+r^2(d\theta^{2}+\sin^{2}\theta d\phi^{2}).
\end{equation}
The singularity at $r=2M$, $x=0$ is removed if $\tau$  is a periodic coordinate with period $8\pi M$. Assuming that the imaginary time 
coordinate $\tau$ is periodic, we obtain a singularity free positive definite Euclidean metric. In fact, the manifold defined by $0\leq x\leq 4M$ and $0\leq \tau\leq 8\pi M$ is the 
Euclidean section of the Schwarzschild solution. 

Let us now consider a $\lambda\varphi^{4}+\rho\varphi^{6}$ scalar model without noise defined in this positive definite Euclidean metric. In order to generate the correlation functions of the model by functional derivatives, as usual, a fictitious source is introduced \cite{Itzy}. The generating functional of the correlation functions of the model is 
\begin{equation}\label{distprob1}
Z(j)=\int_{\partial\Omega} [d\varphi]\,\, \exp\left(-S(\varphi)+  \int d^{4}x\sqrt{g} \,j(x)\varphi(x)\right), 
\end{equation}
where the Euclidean action is 
 $S(\varphi)=S_{0}(\varphi)+S_{I}(\varphi)$. In the above equation $[d\varphi]$ is again a functional measure. The free field effective action $S_{0}(\varphi)$ is given by
\begin{equation}\label{eq:S0}
S_{0}(\varphi)=\int d^{4}x\sqrt{g}\,\frac{1}{2}\varphi(x) \Bigl(-\Delta + m_{0}^{2}\Bigr)\varphi(x)
\end{equation}

\noindent  where $\Delta$ is the Laplace-Beltrami operator in the Euclidean Schwarzschild metric, $g$ is the determinant of the positive definite metric and $S_{I}(\varphi)$ is the self-interacting non-Gaussian contribution, defined by

\begin{equation}
S_{I}(\varphi)= \int d^{4}x\sqrt{g}\,\left(\frac{\lambda_{0}}{4} \,\varphi^{4}(x)+\frac{\rho_{0}}{6}\varphi^{6}(x)\right)\label{10}.
\end{equation}
The symbol $\partial\Omega$ in the functional integral means that the field $\varphi(x)$ satisfies periodic boundary condition in the Euclidean time and we impose Dirichlet boundary conditions for some large radius. In this procedure, surface divergences can appear, however they can be eliminated introducing counterterms as surface interaction \cite{symanzik3,boundary1,boundary2,boundary3}. As we will see, due to the noise effects, there is a sign change in the quartic effective coupling constant. Hence the introduction of the $\varphi^{6}$ term guarantees that there will be a ground state. For a strong noise situation, the absence of such term could lead to a collapse of the system. 

The Laplacian in the Euclidean Schwarzschild coordinates is defined as
\begin{eqnarray}
 \Delta\varphi =-\left(1-\frac{2M}{r}\right)^{-1}\frac{\partial^{2}\varphi}{\partial \tau^{2}}-\frac{1}{r}\frac{\partial}{\partial r}\left(r(r-2M)\frac{\partial\varphi}{\partial r}\right) -\frac{1}{r^{2}}\frac{1}{\sin\theta}\frac{\partial}{\partial \theta}\left(\sin\theta\frac{\partial\varphi}{\partial \theta}\right)\nonumber\\
-\frac{1}{r^{2}}\frac{1}{\sin^{2}\theta}\frac{\partial^{2}\varphi}{\partial \phi^{2}}.
\end{eqnarray}
In order to proceed let us study the free field solution discussed in  \cite{fawcett}. The equation of motion for the free scalar field, $i.e.$ neglecting the non-Gaussian contribution, reads
\begin{equation}
 \Bigl(-\Delta + m_{0}^{2}\Bigr)\varphi(x)=0.
\end{equation}
A standard procedure to solve the above equation consists in using the following separation of variables:
\begin{equation}\label{eq:lm}
\varphi=T(\tau)R(r)Y(\theta,\phi).
\end{equation}
As the imaginary time is a periodic function of period $\beta=8\pi M$, one finds 
\begin{equation}
T_{n}(\tau)=\frac{1}{\sqrt{\beta}}\exp\left(i\frac{2\pi n}{\beta}\tau\right),
\end{equation}
where $n$ is an integer. In  (\ref{eq:lm}), $Y(\theta, \phi)$ will give the well-known spherical harmonics. The radial equation is written as
\begin{equation}
-\frac{1}{r}\frac{d}{dr}\left(r(r-2M)\frac{dR_{nlp}}{dr}\right)-V(r)R_{nlp}=0,
\end{equation}
where 
\begin{equation}
V(r)=\frac{l(l+1)}{r^{2}}+\left(\frac{2\pi n}{\beta}\right)^{2}\frac{r}{r-2M},
\end{equation}
and $p$ refers to the radial eigenvalues. In our case, $p$ is an integer.  Instead of discussing the radial solutions, we are interested to discuss the effects of noise in the Euclidean Schwarzschild manifold. Therefore let us briefly investigate the correlation functions of the model in the absence of the noise. In this situation, the $n$-point correlation functions read
\begin{equation}\label{eq:correlationfunction2}
\langle\varphi(x_{1})..\varphi(x_{k})\rangle=Z^{-1}(j)\frac{1}{(\sqrt{g})^{k}}\left.\frac{\delta^{k} Z(j)}{\delta j(x_{1})...\delta j(x_{k})}\right|_{j=0}, 
\end{equation}
where $\left.Z(j)\right|_{j=0}$ is defined as
\begin{equation}
\left.Z(j)\right|_{j=0}=\int_{\partial\Omega} [d\varphi]\,\, \exp\bigl(-S(\varphi)\bigr). \label{a8}
\end{equation}
These moments of the probability measure are the sum of all diagrams with $k$ external legs, including disconnected ones, with exception of the vacuum diagrams. The generating functional of $n$-point connected correlation functions $W(j)$ is defined as $W(j)=\ln Z(j)$. The order parameter of the model without noise $\langle\varphi(x)\rangle$ is given by
\begin{equation}
\langle\varphi(x)\rangle=Z^{-1}(j)\frac{1}{\sqrt{g}}\left.\frac{\delta Z(j)}{\delta j(x)}\right|_{j=0}. 
\end{equation}

To discuss the effects of random noise we are assuming that is modeled by a zero mean Gaussian white noise. The two-point correlation of the noise is defined by
\begin{equation}
\mathbb{E}[{\delta m^{2}_{0}(x)\delta m^{2}_{0}(y)}]=\sigma\delta^{d}(x-y),
\label{pro111}
\end{equation}
where $\sigma$ is a small parameter that describes the strength of the noise and $\mathbb{E}[\cdots]$ means the average over the ensemble of all the realizations of the noise. This is the situation of a scalar field coupled to a random variable with Gaussian correlations. Since we are interested to investigate the model in the presence of a multiplicative noise, a random contribution $\delta m_{0}^{2}(x)$ is added to $m_{0}^{2}$. In this case the functional action of the  model becomes
\begin{eqnarray}
 S(\varphi,\delta m_{0}^{2}) =\int d^{d}x\sqrt{g}\left[\frac{1}{2}\varphi(x)\Bigl(-\triangle + m_{0}^{2} - \delta m_{0}^{2}(x)\Bigr)\varphi(x)\right.+ \left.\frac{\lambda_{0}}{4}\varphi^{4}(x) + \frac{\rho_{0}}{6}\varphi^{6}(x)\right].\label{eq:hamiltonian}
\end{eqnarray} 

Now, let us study the $n$-point correlation functions associated with the system in the presence of multiplicative noise. The generating functional of correlation functions for one realization of the noise is given by
\begin{equation}\label{eq:generatingdisorder}
  Z(\delta m^{2}_{0};j)= \int_{\partial\Omega} [d\varphi]\,\, \exp\left(-S(\varphi,\delta m_{0}^{2})+\int d^{d}x\sqrt{g} j(x)\varphi(x)\right),
\end{equation}
\noindent where a fictitious source, $j(x)$ as usual is introduced. The $n$-point correlation function for one realization of the noise reads

\begin{equation}
\left\langle \varphi(x_{1})\right.\cdots \left.\varphi(x_{n}) \right\rangle_{\delta m_{0}^{2}}= \frac{1}{Z(\delta m_{0}^{2})}\int [d\varphi]\,\prod_{i=1}^{n}\varphi(x_{i})\exp\left(-S(\varphi,\delta m_{0}^{2})\right),
\end{equation}

\noindent where the noise dependent functional integral that appears in the above equation is defined as  $Z(\delta m^{2}_{0})=Z(\delta m^{2}_{0},j)|_{j=0}$. 
Similarly to a system  without noise, one can define a generating functional for one noise realization, $W_{0}(\delta m^{2}_{0},j)=\ln Z(\delta m^{2}_{0},j)$. Now, we can introduce a noise-averaged correlation function as following
\begin{equation}
\mathbb{E} \left[\left\langle \varphi(x_{1})\cdots\varphi(x_{n}) \right\rangle_{\delta m_{0}^{2}}\right]= \int [d \delta m_{0}^{2}] P\left(\delta m_{0}^{2}\right)\left\langle \varphi(x_{1})\cdots\varphi(x_{n}) \right\rangle_{\delta m_{0}^{2}},
\end{equation}
\noindent where again $[d \delta m_{0}^{2}]$ is a functional measure and the probability distribution of the noise is written as $[d \delta m_{0}^{2}]P\left(\delta m_{0}^{2}\right)$ 
where $P\left(\delta m_{0}^{2}\right)$ is given by

\begin{equation}\label{eq:distprob}
P(\delta m_{0}^{2})=p_{0}\exp\left(-\frac{1}{4\sigma}\int d^{d}x \left(\delta m_{0}^{2}(x)\right)^{2}\right),
\end{equation} 
the quantity $p_{0}$ is a normalization constant. The disorder-averaged generating functional of connected correlation functions is defined as
\begin{equation}
W(j)=\int [d\,\delta m_{0}^{2}]P(\delta m_{0}^{2})\,\ln Z(\delta m_{0}^{2},j).
\label{pro1}
\end{equation}
At this point, the replica method \cite{anderson} can be used to compute the noise average generating functional of connected correlation functions.  Following  \cite{distributional,distributional2,zarro,robinson2}, we are using an alternative approach where the average generating functional of connected correlation functions can be represented by
\begin{equation}\label{eq:completefreeenergy}
W(j)=\Bigg[\sum_{k=1}^{\infty} \frac{(-1)^{k+1}a^{k}}{k!\,k}\mathbb{E}\left[Z^{k}\right] -\log{a} - \gamma - R(a,j)\Bigg],
\end{equation}
where $\gamma$ is the Euler constant, the quantity $R(a,j)$ is given by
\begin{equation}\label{m24}
R(a,j)=-\int [d\,\delta m_{0}^{2}]P(\delta m_{0}^{2})\int_{a}^{\infty}\,\frac{dt}{t}\, e^{-Z(\delta m_{0}^{2},j)t},  
\end{equation}
and finally, $\mathbb{E}\left[Z^{k}\right]$ is the $k$-th integer moments of the generating functional of all correlation functions. The advantage of this approach is that contrary to the replica method this technique neither involves derivatives of the integer moments of the partition function, $\mathbb{E}[Z^{k}]$, nor the extension of this derivative to non-integer values of $k$.
\noindent Notice that $R(a,j)$ vanishes as long as $a\rightarrow\infty$. Indeed, in the following, we discuss the 
asymptotic behavior of $R(a,j)$ which is related to the incomplete Gamma function, defined as \cite{abramowitz}
\begin{equation}
\Gamma(\alpha,x)=\int_{x}^{\infty} e^{-t}t^{\alpha-1}\,dt.
\end{equation} 
\noindent The asymptotic representation for  $|x|\rightarrow\infty$ and $-3\pi/2<\mbox{arg }x<3\pi/2$ reads
\begin{equation}
\Gamma(\alpha,x)\sim x^{\alpha-1}e^{-x}\Big[1+\frac{\alpha-1}{x}+\frac{(\alpha-1)(\alpha-2)}{x^{2}}+\cdots\Big].
\end{equation} 
In the next section we discuss the mean-field approximation and how to go beyond this approximation.

\section{The mean-field approximation of the model with multiplicative noise}\label{sec:vacuum}

As it was shown, the series representation of the average generating functional of connected correlation functions is written in terms of the 
integer moments of the generating functional of all correlation functions $\mathbb{E}\left[Z^{k}\right]$. Using the probability distribution for the noise and the action of the model, this quantity is given by
\begin{equation}\label{eq:ezk0}
\mathbb{E}\left[Z^{k}\right]=\int \prod_{i=1}^{k} \left[ d\varphi_{i}\right] e^{\bigl(-S_{eff}^{\,(1)}\left(\varphi_{i}\right)-S_{eff}^{\,(2)}\left(\varphi_{i},j\right)\bigr)},
\end{equation}
\noindent where the effective actions $S_{eff}^{\,(1)}(\varphi_{i})$ and  $S_{eff}^{\,(2)}\left(\varphi_{i},j\right)$
are written as
\begin{eqnarray}
 S_{eff}^{\,(1)}(\varphi_{i})=\int d^{d}x\,\sqrt{g} \left[\frac{1}{2}\sum_{i=1}^{k}\,\varphi_{i}(x)\Bigl(-\Delta\,+m_{0}^{2}\Bigr)\varphi_{i}(x)\right. +\left.\frac{1}{4}\sum_{i,j=1}^{k}\eta_{ij}\varphi_{i}^{2}(x)\varphi_{j}^{2}(x)\right.\nonumber\\
+\left.\frac{\rho_{0}}{6}\sum_{i=1}^{k}\varphi_{i}^{6}(x)\right],\label{eq:effectivehamiltonian}
\end{eqnarray}
and 
\begin{equation}
S_{eff}^{(2)}(\varphi_{i},j)=\int d^{d}x \sqrt{g}\sum_{i=1}^{k}\varphi_{i}(x)j(x),
\end{equation}
where the symmetric coupling constants $\eta_{ij}$ are given by $\eta_{ij}=(\lambda_{0}\delta_{ij}-\sigma)$. 
The saddle-point equations derived from each integer moment of the generating functional of the correlation functions $\mathbb{E}\left[Z^{k}\right]$ 
in the absence of the external source read
\begin{equation}\label{eq:spe}
\left(-\triangle+m_{0}^{2}\right)\varphi_{i}(x) +\lambda_{0}\varphi_{i}^{3}(x)+ \rho_{0}\varphi_{i}^{5}(x) -\sigma\varphi_{i}(x)\sum_{j=1}^{k}\varphi_{j}^{2}(x) =0.
\end{equation}
Using that $\varphi_{i}(x)=\varphi_{j}(x)$, the above equation becomes
\begin{equation}\label{eq:repsymspe}
\left(-\triangle+m_{0}^{2}\right)\varphi_{i}(x)+\left(\lambda_{0}-k\sigma\right)\varphi^{3}_{i}(x) + \rho_{0}\varphi^{5}_{i}(x) =0.
\end{equation}
Indeed, consider a generic term of the series given by (\ref{eq:completefreeenergy}) with integer moments of the generating functional of the correlation functions
given by $\mathbb{E}\,[{Z^{\,l}}]$. See also Eqs. (\ref{eq:ezk0}) and (\ref{eq:effectivehamiltonian}). For each $k$-th integer moment, $\mathbb{E}\,[{Z^{\,l}}]$, all the fields must be equal, we are led to the following choice in the structure of the fields in each  $\mathbb{E}\,[{Z^{\,l}}]$,
\begin{equation}
\left\{\begin{array}{ll}
        \varphi_{i}^{(l)}(x)&=\varphi^{(l)}(x) \,\,\,\hfill\hbox{for $l=1,2,...,N$}\\
        \varphi_{i}^{(l)}(x)&=0 \quad \,\,\,\,\,\hbox{for $l>N$},
				\end{array}
				\right.\label{RSB1}
\end{equation}

\noindent where for the sake of simplicity we still employ the same notation for the field. Therefore the average generating functional of connected correlation functions is written
as 
\begin{equation}\label{eq:incompletefreeenergy}
 W_{N}(j)=\sum_{k=1}^{N} \frac{(-1)^{k}a^{k+1}}{k!\,k}\mathbb{E}\left[Z^{k}\right]+\cdots.
\end{equation}
In  (\ref{eq:completefreeenergy}), the $W_{N}(j)$ is independent of $a$. However the entire approach relies on the fact $a$ can be chosen large enough so that $R(a)$ can be neglected in practice. In this case, the $W_{N}(j)$ is described by a series which is $a$-dependent. The $a$ factor is incorporated in the functional measure in each integer moment of the generating functional of the correlation functions.

To proceed, the mean-field theory corresponds to a saddle-point approximation in each term of the series. A perturbative approach give us the fluctuation corrections to mean-field theory. With the choice of  (\ref{RSB1}), we get that the integer moments of the generating functional of the correlation functions that defines the average generating functional of connected correlation functions and the  effective action reads
\begin{equation}\label{eq:ezk}
\mathbb{E}\left[Z^{k}\right]=\int \prod_{i=1}^{k} \left[ d\varphi_{i}^{(k)}\right] e^{-S_{eff}^{(1)}\left(\varphi_{i}^{(k)}\right)-S_{eff}^{(2)}\left(\varphi_{i}^{(k)},\, j\right)},
\end{equation}
\noindent where the $S_{eff}^{(1)}\left(\varphi_{i}^{(k)}\right)$ and the $S_{eff}^{(2)}\left(\varphi_{i}^{(k)}\, , j\right)$ are given by
\begin{eqnarray}
 S_{eff}^{(1)}\left(\varphi_{i}^{(k)}\right)=\int d^{d}x\sqrt{g} \sum_{i=1}^{k}\left[\frac{1}{2}\varphi_{i}^{(k)}(x)\left(-\Delta\,+m_{0}^{2}\right)\varphi_{i}^{(k)}(x) + \frac{1}{4}\left(\lambda_{0}-k\sigma\right)\left(\varphi_{i}^{(k)}(x)\right)^{4}\right. \nonumber\\
\left.+\frac{\rho_{0}}{6}\left(\varphi_{i}^{(k)}(x)\right)^{6}\right].\label{eq:effectivehamiltonian2}
\end{eqnarray}
and 
\begin{equation}
S_{eff}^{(2)}(\varphi_{i}^{(k)},j)=\int d^{d}x\sqrt{g} \sum_{i=1}^{k}\varphi_{i}^{(k)}(x)j(x).
\end{equation}
Using  (\ref{RSB1}), we find that
\begin{equation}\label{eq:ezkrsb1}
\mathbb{E}\left[Z^{k}\right]=\mathcal{N}\int  \left[ d\varphi^{(k)}\right] e^{-S_{eff}^{(1)}\left(\varphi^{(k)}\right)-S_{eff}^{(2)}\left(\varphi^{(k)}\right)},
\end{equation}
\noindent where the $S_{eff}^{(1)}\left(\varphi^{(k)}\right)$ and the $S_{eff}^{(2)}\left(\varphi^{(k)}\right)$ are given by 
\begin{eqnarray}
 S_{eff}^{(1)}\left(\varphi^{(k)}\right) =k\int d^{d}x\sqrt{g} \left[\frac{1}{2}\varphi^{(k)}(x)\Bigl(-\Delta\,+m_{0}^{2}\Bigr)\varphi^{(k)}(x) + \frac{1}{4}\bigl(\lambda_{0}-k\sigma\bigr)\left(\varphi^{(k)}(x)\right)^{4}\right.\nonumber\\ 
 \left.+\frac{\rho_{0}}{6}\left(\varphi^{(k)}(x)\right)^{6}\right].\label{eq:effectivehamiltonian2rsb1}
\end{eqnarray}
and 
\begin{equation}
S_{eff}^{(2)}(\varphi^{(k)},j)=k\int d^{d}x\sqrt{g} \varphi^{(k)}(x)j(x).
\end{equation}

The fields in each integer moment of the generating functional of the correlation functions are different since each field has a distinct quartic coefficient $(\lambda_{0}-k\sigma)$. In the series representation for the average generating functional of connected correlation functions, each integer moment of the generating functional of the correlation functions is defined by a functional space where the fields are different.  The average generating functional of connected correlation functions,  (\ref{eq:incompletefreeenergy}) can be written as  
\begin{equation}\label{eq:freeenergya1}
 W_{N}(j)=\sum_{k=1}^{k_{c}} \frac{(-1)^{k+1}}{k!\,k}\mathbb{E}\left[Z^{k}\right]+\sum_{k=k_{c}+1}^{N} \frac{(-1)^{k+1}}{k!\,k}\mathbb{E}\left[Z^{k}\right] \, ,
\end{equation}
\noindent where the first term is the contribution to the average generating functional of connected correlation functions for fields which oscillate around the ground state defined by $\varphi^{(k)}_{0}=0$, for $k\leq k_{c}$. It is possible to go beyond the one-loop approximation using the composite field operator formalism where an infinite of leading diagrams is summed. This technique deals with the effective action formalism for composite operators. One must consider a generalization of the effective action where the scalar field is coupled linearly and quadratically to sources. \cite{jackiw,gino1,gino2,gino3,tarjus}. Going back to our approach, for the latter terms, $i.e.$, $k>k_{c}$, although $\varphi^{(k)}=0$ is a local minimum, another global minimum appears for $|\lambda_{0}-k\sigma|>\frac{4}{\sqrt{3}}m_{0}\sqrt{\rho_{0}}$. A detailed discussion for the ground state structure of the $\rho_{0} \varphi^{6}$ potential is presented in  \cite{drouffe}. 

We shall complete our treatment of the influence of the multiplicative noise in the model discussing the Hawking work \cite{hawkinginteraction} and our new results. Many authors have been discussed the effects of fluctuations in the Hawking temperature. The approach used by these authors is that the mass of the black-hole is a fluctuating parameter \cite{flu1,flu2}. See also \cite{flu3}. Here we are developing an alternative and simpler approach. Using the fact that a spontaneous broken symmetry may be restored at sufficiently high temperature, Hawking claims that a system described by a $\lambda\varphi^{4}$ model can go from a spontaneous broken symmetry phase to the disordered phase near a sufficiently hot black hole. The simplest assumption we can make to model fluctuations in the Hawking temperature is to couple the noise with $\varphi^{2}(x)$. Therefore we have a model for fluctuation in the Hawking temperature. Using the distributional zeta-function approach we obtained a series representation for the average generating functional of connected correlation function. Interpreting each term of the series as quite different subsystems since the order parameter is different in each of them, we have the interesting result that fluctuation in the Hawking temperature is able to generate transition from a disordered states to ordered ones. Therefore for sufficiently strong noise one may expect that the system can goes from a disordered phase to an ordered ones with first-order phase transitions. In the next section, we obtain the $\langle\varphi(x)^{2}\rangle$ in the presence of noise.

\section{The computation of $\langle\varphi(x)^{2}\rangle$ in the presence of multiplicative noise.}\label{sec:instantons}

The aim of this section is to compute the following quantity
\begin{equation}
\langle\varphi(x)^{2}\rangle=\frac{1}{(\sqrt{g})^{2}}\left.\frac{\delta^{2}W_{N}[j]}{\delta j(x)^{2}}\right|_{j=0}.
\end{equation}
From  (\ref{eq:freeenergya1}), the first summation concerns the ground state at $\varphi^{(k)}=0$ and the second refers to the values of $k$ where the ground state occurs at $\varphi^{(k)}\neq 0$. The value of $k_{c}$ represents the $k$ where the ground state goes to $\varphi^{(k)}=0$ to $\varphi^{(k)}\neq0$. From the above discussion, $k_{c}>\frac{\lambda_{0}}{\sigma}+4\frac{m_{0}}{\sigma}\sqrt{\frac{\rho_{0}}{3}}$. Using $\lambda_{{eff}}(k)=\lambda_{0}-k\sigma$. For simplicity, henceforth we are going to use $\lambda_{eff}(k)=\lambda_{eff}$. The field potential is
\begin{equation}
V(\varphi^{(k)})=\frac{1}{2}m_{0}^{2}\bigl(\varphi^{(k)}\bigr)^{2}+\frac{1}{4}\lambda_{{eff}}\bigl(\varphi^{(k)}\bigr)^{4}+\frac{1}{6}\rho_{0}\bigl(\varphi^{(k)}\bigr)^{6}.
\end{equation}
For the case where $k>k_{c}$, the ground state, $\varphi_{0}^{(k)}$, is
\begin{equation}
\varphi_{0}^{(k)}=-\frac{\lambda_{{eff}}}{2\rho_{0}}+\sqrt{\left(\frac{\lambda_{{eff}}}{2\rho_{0}}\right)^{2}-\frac{m_{0}^{2}}{\rho_{0}}}.
\end{equation}

Now we expand the field potential around the new ground state. Defining the field $\varphi^{(k)}=\phi^{(k)}-\varphi_{0}^{(k)}$, we get 
\begin{equation}
V(\phi^{(k)})=V(\varphi_{0}^{(k)})+\frac{m_{{eff}}(k)^{2}}{2}\left(\phi^{(k)}\right)^{2} + \mathcal{O}\left[\left(\phi^{(k)}\right)^{3}\right],
\end{equation} 
where the noise-dependent mass square is
\begin{equation}
m_{{eff}}(k)^{2}=\frac{\lambda_{{eff}}^{2}-4m_{0}^{2}\rho_{0}-\lambda_{{eff}}\sqrt{\lambda_{{eff}}^{2}-4m_{0}^{2}\rho_{0}}}{\rho_{0}}\geq 0.
\label{maeff}
\end{equation}
From now on, for brevity we are going to use $m_{eff}(k)=m_{eff}$. Notice that we retain only the quadratic terms as we are interested to investigate the contributions of the multiplicative noise to the vacuum activity at the Gaussian approximation. Defining a series representation for the vacuum activity, $\langle\varphi^{2}\rangle$, through 
\begin{equation}
\langle\varphi(x)^{2}\rangle = \sum_{k=1}^{N} \frac{(-1)^{k+1}}{k!k}\frac{1}{(\sqrt{g})^{2}} \frac{\delta^{2}}{(\delta j(x))^{2}}\mathbb{E}\left[Z^{k}\right],
\label{renormalizedphi2}
\end{equation}
we get
\begin{equation}\label{eq:rhok}
 \langle\varphi(x)^{2}\rangle=
\sum_{k=1}^{k_{c}} \frac{(-1)^{k}}{(k-1)!}G_{1}^{(k)}(x,x;m_{0})+\sum_{k=k_{c}+1}^{N} \frac{(-1)^{k}}{(k-1)!}G_{2}^{(k)}(x,x;m_{eff})
\end{equation}
where 
\begin{equation}
G_{1}^{(k)}(x,x;m_{0})=\left\langle \varphi^{(k)}(x)\varphi^{(k)}(x) \right\rangle
\end{equation}
is the two-point correlation function associated to the field $\varphi^{(k)}(x)$ which has mass $m_{0}$. In the same way, 
\begin{equation}
G_{2}^{(k)}(x,x;m_{eff})=\left\langle \phi^{(k)}(x)\phi^{(k)}(x) \right\rangle
\end{equation}
is the same quantity for $\phi^{(k)}(x)$ at coincident points, obtained from $k$-th integer moment of the generating functional of correlation functions of the model. The field $\phi^{(k)}$ has mass $m_{eff}$ given by (\ref{maeff}). Therefore we are generalizing the results obtained by Fawcett \cite{fawcett}, computing $\langle\varphi(x)^{2}\rangle$, modified by the presence of multiplicative noise. It is not possible to find the $\langle\varphi(x)^{2}\rangle$ in a closed form. Therefore, in the next section, we are going to consider the region near the event horizon where the Schwarzschild coordinates can be approximated by Rindler coordinates.

\section{The two-point correlation function in Euclidean Rin-\\dler manifold}\label{sec:Rindler}

The aim of this section is use the results obtained in Refs. \cite{duff,c1} to find the $\langle\varphi(x)^{2}\rangle$ in the presence of noise near the horizon. Near the horizon, $r\approx 2M$, taking the Euclidean section ($t\mapsto i4M\tau$) of the Schwarzschild line element (\ref{eq:line element1}), and redefining the radial coordinate as $\rho=\sqrt{8M(r-2M)}$, we get 
\begin{equation}\label{eq:RindlerA}
ds^{2}=\rho^{2}d\tau^{2}+d\rho^{2}+4M^{2}d\Omega^{2},
\end{equation}
where the horizon is located at $\rho=0$ and the angular part of this metric describes a line element of a 2-sphere with radius $2M$. Using Eqs. (\ref{eq:S0}) and (\ref{eq:RindlerA}), the equation of motion in the Euclidean Rindler space is given by
\begin{equation}
\left(\frac{1}{\rho^{2}}\partial_{\tau}^{2}+\partial_{\rho}^{2}+\frac{1}{\rho}\partial_{\rho}+\partial_{i}^{2}+m_{0}^{2}\right)\varphi=0,
\end{equation}
which normalized\footnote{The inner product is $\langle\varphi,\psi\rangle=\int_{M}\varphi^{*}(x)\psi(x)g^{\tau\tau}\sqrt{g}d^{d-1}x$.} modes can be obtained
\begin{equation}
u_{\vec{k}_{\perp},\omega}(\tau,\rho,\vec{x}_{\perp})=\frac{\sqrt{2\omega\sinh\pi\omega}}{\pi(2\pi)^{\frac{d-2}{2}}}K_{i\omega}(\mu\rho)\exp(i\vec{k}_{\perp}\cdot\vec{x}_{\perp}+\omega\tau),
\end{equation}
where $\mu=\sqrt{\vec{k}^{2}_{\perp}+m_{0}^{2}}$ and $K_{i\omega}(x)$ is the modified Bessel function of third kind. From these modes, we get the two-point correlation function \cite{c1,maria1,maria2,linet}:
\begin{eqnarray}
 G_{2\pi}(\tau,\rho,\vec{x}_{\perp},\tau',\rho',\vec{x}^{\prime}_{\perp})=\int_{0}^{\infty}d\omega\;\int d^{d-2}\vec{k}_{\perp}\frac{\sinh\pi\omega}{\pi^{2}(2\pi)^{d-2}}\nonumber\\
\times K_{i\omega}(\mu\rho)K_{i\omega}(\mu\rho')\exp\left(i\vec{k}_{\perp}\cdot(\vec{x}_{\perp}-\vec{x}^{\prime}_{\perp})\right)\exp\left(-\omega|\tau-\tau'|\right).
\end{eqnarray}
From the (\ref{eq:rhok}) we have that $\langle\varphi(x)^{2}\rangle$ has two kind of contributions constructed with $G_{1}(x, x)$ and $G_{2}(x,x)$. Since both contributions are divergent, to regularize this divergent quantities we are following Refs. \cite{maria1,maria2,linet}. We define the renormalized two-point correlation function of coincident points as
\begin{equation}
\left[G_{i}^{(k)}(x,x)\right]_{{ren}}=\left.\left(G_{2 \pi}^{(k)}\left(x, x_{0} ; m_{i}\right)-G_{\infty}^{(k)}\left(x, x_{0} ; m_{i}\right)\right)\right|_{x=x_{0}},
\end{equation}
where $G_{2 \pi}^{(k)}\left(x, x_{0} ; m_{i}\right)$ is the finite temperature ($T=1/2\pi$)-Rindler Schwinger function\footnote{It is shown in \cite{c1} that this very Schwinger function is exactly the Euclidean one.}   and $G_{\infty}^{(k)}\left(x, x_{0} ; m_{i}\right)$ is the ($T=0$)-Rindler-Schwinger function. For simplicity, we consider the four dimensional case. The mass $m_{i}$ can be either $m_{0}$, for $i=1$, or $m_{eff}$, for $i=2$, where $m_{eff}^2$ is defined in (\ref{maeff}). The above quantity can be written as \cite{linet}: 
   \begin{equation}\label{eq:renormalizedSchwinger}
        \left[G_{i}^{(k)}(x,x)\right]_{{ren}}=\frac{m_{i}}{4 \pi^{2} \rho} \int_{0}^{\infty} d u \frac{K_{1}[2 m_{i} \rho \cosh (u / 2)]}{\left(\pi^{2}+u^{2}\right)\cosh (u / 2)}.
    \end{equation}
Finally, at the tree level, using Eqs. (\ref{eq:rhok}) and (\ref{eq:renormalizedSchwinger}), we get the renormalized expectation value of the scalar field $\varphi^{2}$ in the presence of noise:
\begin{equation}\label{eq:vactivity}
 \langle\varphi(x)^{2}\rangle_{{ren}}=
\left(\sum_{k=1}^{k_{c}} \frac{(-1)^{k}}{(k-1)!}\right)\left[G_{1}^{(k)}(x,x)\right]_{{ren}}+\sum_{k=k_{c}+1}^{N} \frac{(-1)^{k}}{(k-1)!}\left[G_{2}^{(k)}(x,x)\right]_{{ren}}.
\end{equation}
We obtained the ``vacuum activity'' near the event horizon associated to a scalar field in the presence of multiplicative noise in the Gaussian approximation. Now we are going to interpret our results. In the absence of the $\varphi^{6}$ term, there is a possibility of the collapse of the system since the effective coupling constant $\lambda_{eff}=\lambda_{0}-k\sigma$ becomes negative for a sufficiently strong noise. With the presence of the $\varphi^{6}$ term, he have now two possibilities: 
 the vacuum activity is represented only by the first sum of the right hand side of  (\ref{eq:vactivity}) or by the whole series. In the former, the noise is so weak that all fields have the same mass $m_{0}$. In the latter, for $k>k_{c}$, the vacuum activity is represented by a series of correlation functions of fields of different effective masses, $m_{eff}$. 

Now, to conclude, we are going to take the limit where the non-random mass is zero, $m_{0}=0$. In this case, some exact results can be obtained. Equation (\ref{eq:renormalizedSchwinger}) can be simplified (see Ref. \cite{linet})
\begin{equation}
\left[G_{1}^{(k)}(x,x)\right]_{{ren}}=\frac{1}{8\pi^{2}\rho^{2}} \int_{0}^{\infty} d u \frac{1}{\left(\pi^{2}+u^{2}\right)\cosh^{2} (u / 2)}=\frac{1}{48\pi^{2}\rho^{2}}.
\end{equation}
Using  (\ref{maeff}) we have $m_{eff}=0$ and $\left[G_{1}^{(k)}(x,x)\right]_{{ren}}=\left[G_{2}^{(k)}(x,x)\right]_{{ren}}$. The vacuum activity for a massless scalar field can be computed
\begin{equation}\label{eq:vactivitym0}
 \langle\varphi(x)^{2}\rangle_{{ren}}= \left(\sum_{k=1}^{N} \frac{(-1)^{k}}{(k-1)!}\right)\frac{1}{48\pi^{2}\rho^{2}}.
\end{equation}
This series converges very fast and  the vacuum activity is written as
\begin{equation}\label{eq:vactivitymassless}
 \langle\varphi(x)^{2}\rangle_{{ren}}= -\frac{1}{48 e \pi^{2}\rho^{2}}.
\end{equation}
Therefore the presence of noise changes very abruptly the nature of the vacuum activity of a scalar field near the event horizon of a Euclidean Schwarzschild black hole. This quantity turns to be negative if the noise is present rather than the positive one for the clean system. For $m_{0}\neq 0$, one has to perform numerical calculations since the integral of  (\ref{eq:renormalizedSchwinger}) cannot be solved in a closed form. However, we see that this integral is always positive, and the the effect of noise is to change the vacuum activity to a negative value no matter how weak is the disorder strength.

\section{Conclusions}\label{conclusions}

Nonlinear stochastic differential equations have been used to describe different systems from nonlinear quantum optics, non-equilibrium growing interface to growth mechanism of population dynamics.  On general grounds, stochastic differential equations are usually defined with additive or multiplicative noise. The linear stochastic differential equation with multiplicative noise has interesting consequences in the physics of nonequilibrium systems. 

Inspired in the above scenario, we discuss a $\lambda\varphi^{4}+\rho\varphi^{6}$ scalar model defined in the Euclidean section of the Schwarzschild solution of the Einstein equation in the presence of multiplicative noise. The multiplicative noise models a fluctuating Hawking temperature. Before continuing, we would like to stress that our approach is much less ambitious than the stochastic semiclassical gravity where is include fluctuations in Einstein equations, defining the stochastic semiclassical Einstein Langevin equations \cite{ech1,ech2}. See also \cite{sg}. The idea of this approach is to solve self-consistently these equations discussing back-reaction of the quantum stress-energy fluctuations of the gravitational field. In our approach the back-reaction problem is ignored.

We adopt the standard procedure of averaging the generating functional of connected correlation functions of the model in the presence of the noise. The dominant contribution to this quantity is represented by a series of the moments of the generating functional of correlation functions of the system. In the Euclidean functional approach it is necessary to define a disordered functional integral of the system using the disordered action. From the disorder generating functional for one realization of the noise of the model we must obtain the  noise averaged generating functional of the model. To obtain the quantity discussed before we will use the distributional zeta-function method. We obtain a series representation for the generating functional of connected correlation functions. Some terms of the series that contribute to the average generating functional of connected correlation functions is defined for fields which oscillate around the ground states defined by $\varphi^{(k)}_{0}=0$. This series representation contains new interesting features. For $k>k_{c}$, different global minimum in each term of the series appears. To proceed we computed, in the Gaussian approximation, $\langle\varphi(x)^{2}\rangle$ in the presence of noise near the horizon. An exact result can be obtained for $m_{0}=0$ and the presence of noise changes completely the behavior of the scalar field in a Euclidean black hole.

In this article we have discussed an unified mathematical description of noise-induced phase transitions. A natural continuation of this work is the following. From  (\ref{eq:freeenergya1}) it is possible to compute the two-point correlation function $G(x,x')$.  Then we analytically continue it to the Lorentzian metric to obtain the positive frequency Wightman function. The Fourier transform of this quantity will give the response function of a Unruh-Dewitt detector interacting with the scalar field with the presence of multiplicative noise.  Another related issue concerns the possibility to construct an analog model for the fluctuating Hawking temperature. These subjects are under investigation by the authors.

\section*{Acknowlegements} 
We would like to thank G. Krein and  G. Menezes for useful discussions. This work was partially supported by Conselho Nacional de Desenvolvimento Cient\'{\i}fico e Tecnol\'{o}gico - CNPq, 309982/2018-9 (C.A.D.Z.) and 303436/2015-8 (N.F.S.).

\end{document}